# Quantum Universe and its Elusive Classicality

Jahan N. Schad PhD., Retired LBNL (UCB) Scientist, USA\

Schadn5@Berkely.edu

The article delves into the intricate challenges presented by recent revelations in physics, particularly within the domain of quantum mechanics, and the profound questions they provoke concerning our understanding of reality. It sheds light on the disparity between the diffuse, indefinite nature of quantum reality—the bedrock of existence—and our perception of a singular classical reality. The necessity for a universal transition or reduction from the quantum to classical reality is underscored, alongside the hurdles associated with observer intervention and the complexities introduced by the objective collapse theory. It acknowledges the interpretational challenges of quantum-to-physical reduction, coupled with the dependence on ad hoc collapse theories. The work introduces a pioneering approach utilizing the concept of decoherence within the framework of universal wave function density matrices. It centers on the entanglement among three primary quantum subsystems: mass particles, massless particles, and the human body-nervous system. This complex entanglement gives rise to a "Von Neumann chain" of correlated systems, wherein neglecting one system renders the other two entangle subsystems in mixed states. Each mixed state comprises statistically correlated objective indefinite alternatives. In the proposed approach, the "two mixed quantum subsystems" under scrutiny—the cohered "massless (mainly photonic) system" and the "animate being"—constitute subjects of our perceptions. The probabilistic alternative states of the body ensemble can be construed as microstates in the realm of statistical mechanics. Within the body subsystem, the distributions of particle wave packets are highly localized due to entanglement and chemical potential, satisfying the Ehrenfest condition—wherein the thermal deBroglie wavelength scale is considerably smaller than the domain defining the particle distribution within the body. This facilitates the consideration of the movement of packet centers according to classical mechanics. Thus, each microstate of the body particle ensemble in quantum mechanics can be analogously represented as an appropriate statistical mechanics particle ensemble microstate (e.g., microcanonical) or more realistically likened to microstates in the Debye statistical model, where microstates represent different vibration modes of the ensemble. Life inherently embodies one of the microstates of such an ensemble—potentially it may have originated from a quantum microensemble. In this statistical mechanics perspective, life represents "a macrostate" characterized by thermodynamics parameters such as entropy, internal energy, and chemical potential. The life macrostate autonomously processes an environment (via the nervous system machinery) based on a corresponding indefinite mixed state alternative of its entangled photon and phonon quantum world (subsystem) counterpart, which subsequently manifest in the stream of consciousness. The quantum universe persists, – the universal wave function does not collapse, evolving deterministically— and the classical (macroscopic) universe is its statistical realization by the fundamental predicate of life discerned by the brain; our elusive world.

Introduction

Scientific theory posits that the initial stages of the universe, preceding the Big Bang [1; 2], led to the emergence of its quantum fields and primordial plasma composed of quantum particles. This plasma, a blend of interacting fields, comprised primarily of photons and quarks [3]. From this inception, as far as current understanding indicates, the fundamentally (microscopically) quantum mechanical universe with the physical (classical) appearance we perceive, has emerged. It is worth noting that the question regarding the reality of the world we perceive has long intrigued our predecessors throughout history. This inquiry finds resonance in Plato's concept of the "other world" and his allegory of the cave [4; 5], as well as in philosopher Schopenhauer's contemplation of the "thing in itself [6]." Additionally, some contemporary philosophers have proposed the intriguing notion that the universe could potentially be akin to a computer simulation, thus a virtual reality [7; 8]. It is somewhat bewildering that the inherently quantum nature of the universe also gives rise to such uncertainties. This is due to the peculiar nature of its constituent particles. Quantum particles exhibit a multimodal, indefinite objective existence, possessing different characteristics and occupying various locations, all with a certain probability [9]. These states are represented by a wave (packet) function [10], which evolves according to Schrödinger's wave equation, reflecting the probabilistic nature of quantum mechanics. Despite the captivating insights into the enigmatic workings of nature and the multitude of evidences supporting quantum mechanics theory, it has thus far fallen short in elucidating the "quantum-to-physical process." This process, seemingly responsible for the transition to physicality (classicality) experienced across all levels, from the empirical validations of its theoretical foundations in laboratories to the broader scope of the universe, remains a puzzle yet to be fully unraveled. There are also a few other misgivings in this field, essentially the problem of non-locality (violation of special relativity) and integration with the theory of relativity. Physicist Paul Davies, in a recent presentation [11], categorizes all concerns of the field under the topic of "open question (minutely rephrased) of quantum mechanics" as follows:

1) *Is everything quantum?*
2) *Does quantum mechanics have a restricted range?*
3) *Is quantum mechanics an effective theory to be replaced?*

Given the robustness of quantum theory, substantiated by numerous experimental observations and practical applications in quantum physics, it is highly probable that future discoveries will further complement and enrich our understanding of this foundational framework. Regarding the puzzle of the quantum to-physical transition (reduction) process," the first major idea has been the Copenhagen interpretation [12]. This theory primarily concerns the interpretation of measurements of microscopic quantum systems by an observer. In quantum terminology, this is described as the collapse of the quantum particle's wave function from its objectively indefinite state into a definite objective state. This approach suggests that the boundary between quantum and physical reality extends to the mind of the experimenters [13], empowering them to investigate quantum reality and its perceived classicality. The interpretation, aimed at resolving what was referred to as the "measurement problem," encountered skepticism within the physics community, as reflected in the aphorism "shut up and calculate."The Copenhagen interpretation was mathematically formalized by the well know physicist Von Neumann [14], where a procedure called" process 1" accounts for the reduction. And the theorized process purports a link between two realms; the mind (consciousness) and physicality, which is behind the reduction process. The role of the consciousness in the quantum to-physical reduction was also alluded to by the Nobel laureate Physicist Wigner [15] in the following statement:

*"that the consciousness of an observer is the demarcation line that precipitates collapse of the wave function, independent of any realist interpretation."*

Physicist Stapp [16], in reviewing earlier works and emphasizing the role of the consciousness, points to its elusive nature in the following statement, which nonetheless imparts a hint of doubt about the autonomous mind (an agency that even its denier have an impression of having):

*"At the pragmatic level it is a "free choice," because it is controlled, at least in practice, by the conscious intentions of the experimenter/participant, and neither the Copenhagen nor von Neumann formulations provide any description of the causal origins of this choice, apart from the mental intentions of the human agent."*

In a further elaboration on Von Neumann's formulation, Stapp elucidates the dynamics of process 1 within the realm of ontology, akin to Descartes' psycho-physical dualism-- a belief that to some extent, suggests that the "sense of autonomy" maintained by the mind finds its counterpart in the brain, enabling beings to perceive the ability to choose. A modified version of orthodox quantum mechanics, known as the Von Neumann/Stapp approach [17], incorporates additional processes such as "process 3," wherein Nature accommodates the choices contemplated in process 1. Nevertheless, the concept of "choice" inherently pertains to consciousness and autonomous mind. However, prevailing consensus in contemporary philosophy, alongside modern cognitive and hard sciences, suggests skepticism towards the notion of free will, viewing consciousness in a different light. This sentiment is succinctly captured in a statement by Zurek [18], particularly in the context of the realization or transition from quantum reality to classicality.

*"Moreover, while the ultimate evidence for the choice of one alternative resides in our elusive 'consciousness,' there is every indication that the choice occurs much before consciousness ever gets involved and that once made, once made, is irrevocable."*

As evident from the statement, consciousness appears to be positioned as the recipient of choices generated by the brain, implying that choices are primarily physical (neural) phenomena occurring within the brain before being translated into the stream of consciousness. The works of Libet [19] and similar studies likely underpin this interpretation. Libet's research suggests that choices are essentially "physical workings of the brain." However, Stapp [17] addresses this notion within the framework of quantum mechanics formalism, proposing that the quantum state of the brain is influenced by the sustained (rapid) "intentions of the agent" and the "neural correlates of the brain," which inevitably acknowledges the two processes ascribed to consciousness. Stapp's work represents an elegant and perhaps tactful endeavor to incorporate the role of the "willful agent" into the process of quantum state measurements by establishing a connection between the realms of mind and brain, as intentionality persists. "Consciousness," arising from our brain's operations, is commonly perceived as the enigmatic aspect of our nervous system; it constitutes the autonomous mind that we all have impression of having it. Philosophers spanning millennia and cognitive sciences in recent decades have grappled with gaining insights into its nature, yet it remains predominantly a "hard problem [20]." However, in certain circles of thought, it is non-issue and could be elucidated within the framework of system theory [21]. This perspective views "consciousness as the output of the nervous system (intelligent computing machinery) in response to sensory stimuli," achieved through the activation of motor neurons in various ways; embodiment of consciousness [22]. This rationale, though unconventional, aligns with the notion that every event occurring within a system can be interpreted as the outcome of computations governed by physical laws, system specifications, and the nature of stimulation from physical processes. The significant role of consciousness also emerges in the advance reduction theory of "Decoherence" authored the physicist Peter Zeh [23] as pointed out below:

*"Accordingly, it is the observer who "splits" indeterministically—not the (quantum) world."*

Decoherence theory attributes quantum wave function collapse to the loss of coherence as a result of entanglement with the measurement environment. Physicist Zurek in his elegant paper regarding the subject [18] demonstrated the process very clearly in an example. These works [23; 18] have provided a basis for the understanding of some of the fundamental aspects of the "quantum-to physical transitions," and possibly even the second law of thermodynamics.

Given the mystical aspects of collapse theories, a critical appraisal is imperative: If "the physical world is real," asserting that life had anything to do with it ventures beyond the scope of science. Therefore, the quest for its emergence must explore avenues such as the "Everettian approach" or the "objective collapse theories." Everett pursued the concept of classicality within the context of the universal wave function, stemming from the universe's early quantum origins. DeWitt [24] initially proposed the notion of the "universal wave function," extensively pursued by Everett [25]. The approach entails an infinitude of universes, which presents challenges in reconciling with our classical universe, especially considering the closed system status of the quantum universe, and absent external observer. The objective collapse theories show promising prospects, as initially proposed by one of the pioneers of quantum mechanics, as cited by Stapp [16], in the following:

*"---- Heisenberg introduced the Aristotelian concept of "potentia", and regarded the quantum mechanical state of a system to be not only a compendium of knowledge about what has happened in the past, but also a "potentia"---an objective tendency- --for this evolving quantum*

*state to abruptly collapse to a reduced part of itself. These reductions are needed to keep cutting back the otherwise expanding continuum of possibilities created by the Schrodinger-equation-based temporal evolution of the quantum state to the part of itself that is compatible with our collective human experience."*

This philosophical assertion posits that autonomous reduction is inherent in the nature of the quantum state of the universe. The objective collapse theories have issues concerning the conservation of energy and the special theory of relativity. Nevertheless, ongoing research in both domains holds promise in addressing concerns and aligning with scientific principles. While physicists' remarkable efforts to consolidate a robust theoretical basis for the quantum-to-physics process continue, it may also be prudent to consider the possibility of reduction being merely a phenomenon of the brain, drawing upon some aspects of the early approaches of the founders of quantum mechanics: it is conceivable that the classical world, in one form or another, is a construct of the human brain encapsulating the quantum wave function of its constituent particles. This possibility becomes more tangible if we note that our perceived classical universe, in the final analysis, is a simulation crafted by the brain, interpreting external stimuli through electrochemical signals and computational processes. This perspective reflects a deterministic viewpoint [26; 21; 22], attributing causality to all brain dynamics and dismissing the notion of autonomous mind; thus, the perceived classicality of the universe results from computations within our autonomous nervous system. The presented work offers a non-collapse hypothesis of reduction that maintains the universe in a quantum state while still explaining the perceived classicality, without recourse to consciousness or the autonomous mind. This approach, known as "quantum-physical dualism," replaces the traditional "Psycho-Physical" dualism, enabling the evolution of our universal quantum existence despite experiencing it in a classical manner. Moreover, it would address concerns raised by Stapp [16] regarding the potential survival of human personality beyond bodily death.

The Hypothesis

It is common assumption in the realm of physics that our perceived universe, in which we participate, adheres to classical physical principles and has emerged from a primordial quantum plasma entity. Despite its classical appearance, this familiar world has persisted, at its core, as fundamentally (microscopically) quantum, as elucidated by the Standard Model of elementary particles. This theoretical framework of the quantum universe, as expounded by Von Neumann [14], serves as our starting point:

*"...quantum theory is a formulation in which the entire physical universe, including the bodies and brains of the conscious human participant/observers, is represented in the basic quantum state, which is called the state of the universe."*

The quantum state of the universe can be conceptualized within the framework of a hypothetical relativistic universal wave function, grounded in the quantum inception of the universe, governing its evolutionary trajectory. This function, housing numerous objective indefinite alternatives, serves as a hypothetical solution to the foundational law of quantum mechanics, Schrödinger's equation, enabling the definition of wave packet distributions for every particle in the universe. Addressing the genesis of the classical world through the lens of a universal quantum wave function presents opportunities for resolving the inherent puzzle of reduction unburdened by the shortcoming of the quantum wave function collapse approach. An important aspect of the formalism of the universal wave function is to recognize that the quantum system of the universe likely comprised an infinite array of clusters early on due to entanglements and interactions among particles, shaping the distribution of masses (animate and otherwise) observed in the universe. This suggests that the universal wave function would encompass infinity of objective indefinite alternatives with varying probabilities within mass clusters, while maintaining nearly equally probable distributions for non-interacting massless particles within their domain. This condition finds optimal representation in the formulation of density matrices—an alternative means of representing the probabilistic states of quantum systems—for the entire system or its various segments. Building upon this conceptual foundation, I elucidate the classicality of the universe we perceive within the context of density matrices pertaining to the tri-segment (subsystems) partition of the "pure quantum system of the universe," forming a "Von Neumann chain" of correlated systems. Disregarding (tracing over, in mathematical terminology) the density of one exceedingly complex segment results in "mixed states" for the remaining two, each exhibiting "statistical correlation"—an outcome akin to applying "Von Neumann's non-unitary process 1" between two subsystems. In the proposed approach, the "two mixed quantum subsystems" under scrutiny—the cohered "massless (mainly photonic) system" and the "animate being"—constitute subjects of our perceptions. The probabilistic alternative states of the body ensemble can be construed as microstates in the realm of statistical mechanics. Within the body subsystem, the distributions of particle wave packets are highly localized due to entanglement and chemical potential, satisfying the Ehrenfest condition—wherein the thermal deBroglie wavelength scale is considerably smaller than the domain defining the particle distribution within the body. This facilitates the consideration of the movement of packet centers according to classical mechanics. Thus, each microstate of the body particle ensemble in quantum mechanics can be analogously represented as an appropriate statistical mechanics particle ensemble microstate (e.g., microcanonical) or perhaps more realistically likened to microstates in the Debye statistical model, where microstates represent different vibration modes of the ensemble. Life inherently embodies one of the microstates of such an ensemble—potentially it may have originated, from certain ensemble of quantum particles. This possibility is bolstered by Schrödinger's proposition, further elaborated upon by the esteemed physicist Paul Davies, suggesting that quantum mechanics could offer insights into the mystery of life's origins. [27]. In this statistical mechanic realization of the quantum ensemble state of the body, life represents "a macrostate" characterized by thermodynamics parameters such as entropy, internal energy, and chemical potential. The life macrostate autonomously processes an environment (via the nervous system machinery) based on a corresponding indefinite mixed state alternative of its entangled photon and phonon quantum world (subsystem) counterpart, which subsequently manifest in the stream of consciousness-- it is crucial to note that photons and phonons are carriers of information about the rest of the quantum universe. Essentially, our nervous system primarily interacts with photons and phonons, entangling with them in a coherent manner within the vast quantum realm. This interaction stimulates our senses and plays a role in shaping classical perception. From a heuristic standpoint, one might propose that our universe is essentially a quantum photonic segment permeating the seemingly empty space— the realm perceived by our senses [28]. Our direct tactile sensation, in particular, can be attributed to the influence of phonons.

In the framework of the aforementioned thesis, the quantum universe persists, – the universal wave function does not collapse, evolving deterministically— and the classical (macroscopic) universe is its statistical realization by the fundamental predicate of life discerned computationally by the brain. The classic computed world could perhaps be thought of as the ensemble realization of the concept of the Pointer State [18]. In final analysis I concur with Peter Zeh [23]: *it is the observer who "splits indeterministically – not the (quantum) world.*

Conclusion

The provided passage delves into the physics challenge of comprehending the shift from the early post-inflationary universe, characterized by plasma of quantum particles, to the classical (macroscopic) reality we perceive. This transition poses an unresolved puzzle in physics, offering two potential avenues for resolution: satisfying the need for an observing external entity, absent as far as known, or uncovering an autonomous process to facilitate the transition. However, both of these approaches hinge on the "direct reduction" of the quantum universe, either through objective collapse or via autonomous minds, both of which encounter significant hurdles and fail to fully elucidate the quantum reality experienced at the Macroscopic , or even microscopic levels. This work, inspired by the concept of decoherence, suggests a hypothesis that leaves the quantum reality of the world intact, which may also help resolving some of quantum mechanics shortcomings. The hypothesis involves considering three coherent quantum subsystems that collectively form the entire pure quantum system of the universe. These subsystems include the animate quantum medium (beings' nervous system, and body), the massless quantum medium (mainly photons and phonons), and the external quantum mass particle system of the universe. This division, forming a "Von Neumann's chain [18]," allows consideration of the density matrices of the first two subsystems in mixed state as a result of ignoring (tracing over) that of the third one. This operation results in the body-nervous system and the photonic-photon systems becoming statistically correlated (mixed quantum) subsystems systems, which could be treated in the realm of statistical mechanics, where a macrostate defining life computes the classical conception of the environment, a world in which it can sustain itself.

Acknowledgement

This work is inspired by a publication of Zurek [18], which addresses the eventuality of triply entangled simple quantum system, where the density formalism of the quantum states of these systems allows for the development of mixed state for two, by ignoring the third, without resort to Von Neumann's process 1.
References

1- Menzel, David (1932). "A blast of Giant Atom created our universe". *Popular Science*. Bonnier Corporation. p. 52.
2- Guth Alan (2002). "Inflation and the New Era of High-Precision Cosmology". physics@mit. MIT Department of Physics.
3- Press, W. H. (1990). "The Early Universe. Edward W. Kolb and Michael S. Turner. Addison-Wesley, 547 pp., illus. Frontiers in Physics, 69". *Science*. **249** (4970):808–809.
4- Penrose R. (1990). Emperor's New Mind. Oxford England
5- Gail F. (1990). Plato; Metaphysics and Epistemology, Oxford, England
6- Janaway, Christopher. (1999). "Knowing the Thing in Itself." In Self and World in Schopenhauer's Philosophy, 188-207. Oxford: Oxford University Press
7- Bostrom N. (2003) Are You Living In a Computer Simulation?" Philosophical Quarterly ;53(211):243–255
8- Schad J. N. (2023). Mathematics and Physics; Brain and Universe Machine Languages, J. Neurol Stroke. 13(1):4–7
9- Born, M. (1970). *The Republic*. Oxford University Press
10- de Broglie, L. (1925). On the theory of quanta, Dissertation. Translated in 2004 by A. F. Kracklauer.
11- Davies, P. (2019). Open Questions at the Quantum Frontier. Institute for Quantum Studies. https://www.youtube.com/watch?v=2D-F0Wfu98Q
12- Bell J. S. (2011). "Speakable and Unspeakable in Quantum Mechanics," Cambridge University Press, 2nd edition, 2004. Online publication
13- Hossenfelder S. Consciousness and Quantum Mechanics; how are they related? https://www.youtube.com/watch?v=v1wqUCATYUA
14- Von Neumann J. (1955). Mathematical Foundations of Quantum Mechanics. (Translated by R. T. Beyer) Princeton University Press
15- Wigner, Eugene; Henry Margenau (1967). "Remarks on the Mind Body Question, in Symmetries and Reflections, Scientific Essays". *American Journal of Physics*. **35** (12): 1169–1170
16- Stapp, H. P. (2000). From Quantum Nonlocality to Mind-Brain Interaction. LBNL-44712. Lawrence Berkeley National Laboratory, University of California